\documentstyle[12pt]{article}
\begin{document}
\thispagestyle{empty}
\begin{titlepage}
\title{Large order asymptotics of semiclassical expansion: a new approach}

\author{
   O.Yu. Shvedov  \\
{\small{\em Institute for Nuclear Research of the Russian
Academy of Sciences,  }}\\ {\small{\em 60-th October Anniversary prospect 7a,
Moscow 117312, Russia
}}\\    {\small and}\\ {\small {\em Moscow State University,}}\\
{\small{\em Vorobievy gory, Moscow 119899, Russia}}   }

\end{titlepage}
\maketitle

\begin{center}
{\bf Abstract}
\end{center}

A new approach to the problem of finding
 the  asymptotical behaviour of large orders of
semiclassical expansion is suggested. Asymptotics of high orders not only
for eigenvalues, but also for eigenfunctions, are constructed. Thus, one
can apply not only functional integral technique, which has been used up to
now, but also method of direct analysis of the
semiclassical expansion recursive relations.

\newpage

\section*{1 Introduction}

Problems associated with  large order behaviour of perturbation series
coefficients were examined in many papers. It was F.Dyson \cite{D} who
argued in 1952 that perturbation series in quantum field theory diverges.
Asymptotics of coefficients at high orders which is useful, for example,
for investigation of the divergent series summation problem was found later.

This asymptotics is usually constructed by the technique used by L.Lipatov
\cite{L} in quantum field theory and by E.Br\'{e}zin, J.C.Le Guillou and
J.Zinn-Justin \cite{BLGZJ} in quantum mechanics. This method is the following.
The  $k$-th order of perturbation theory for quantities like Green functions
can be represented through a functional integral which can be approximately
calculated by saddle-point technique. This approach was reusable in quantum
mechanics (see, for example, \cite{BLGZJ,BPZJ,ZJ,RS}) for study of the
high orders of perturbation theory for ground state energy and the $n$-th
excited state energy as $n$ is not large.

In this paper a new method for constructing high order asymptotics is
suggested. This technique allows us to find such asymptotics not only
for eigenvalues but also for eigenfunctions. This method is based not on the
functional integral approach but on direct analysis of recursive relations.

Consider the following dependence of the Hamiltonian on a perturbation
theory parameter $g$, momenta $p=(p_1,...,p_n)$ and coordinates
 $x=(x_1,...,x_n)$:
 \begin{equation}\label{11*}
 {\cal H}=\frac{p^2}{2}+\frac{1}{g^2}V(gx),
 \end{equation}
where $p_m=-i\partial/\partial x_m$, the potential $V$ has a local minimum
at $x=0$. These Hamiltonians were considered in \cite{BLGZJ,BPZJ,ZJ}. Without
loss of generality, one can assume that
$V(Q) \sim Q^2/2 + O(Q^3)$
as $Q \rightarrow 0$.

Contrary to the perturbation expansion in powers of $g$
for eigenvalues, perturbation expansion for eigenfunctions can be constructed
in different ways. For instance, the wave function $\Psi$ can be simply
expanded in powers of $g$ at fixed $x$,
$\Psi(x)=\sum_{k=0}^{\infty} g^k\Psi_k(x)$.
But on the other hand, one can first carry out the following change of the
wave function argument,
 \begin{equation}\label{11+}
 gx=Q,
 \end{equation}
and expand the wave function at fixed $Q$. It appears that in this case the
expansion in powers of $g^2$ is a tunnel semiclassical expansion (see, for
example, \cite{LL,M1}). The square of perturbation theory parameter becomes
the Planck constant analog, because the equation for eigenfunction
$ {\cal H}\Psi = E_0 \Psi$
after multiplying it through by $g^2$, substitution (\ref{11+})
and renotation $g^2=\hbar$ takes the form
\begin{equation}
-\frac{\hbar^2}{2}\Delta\Psi + V(Q) \Psi
= \hbar E_0 \Psi.
\label{*3}
\end{equation}

As it is known (see, for example, \cite{LL,M1}), the tunnel asymptotics of the
ground state wave function has the form of a product of a slowly varying
pre-exponential factor by a rapidly varying exponentional function
\begin{equation}
\Phi(Q,\hbar)\exp\left(-\frac{1}{\hbar}S(Q)\right).
\label{D1*}
\end{equation}
Substitution of this formula to eq.(\ref{*3}) gives the Hamilton-Jacobi
equation for function $S$
\begin{equation}
(\nabla S(Q))^2/2=V(Q)
\label{*1}
\end{equation}
and the following equation for function  $\Phi$,
\begin{equation}
-\frac{\hbar}{2}\Delta\Phi + \nabla S \nabla \Phi +
\frac{1}{2}(\Delta S - n)\Phi= (E_0-n/2)\Phi.
\label{D2*}
\end{equation}
One can apply perturbation theory to it. Both function $\Phi$ and ground
state energy can be expanded in powers of $\hbar$,
$$
\Phi(Q,\hbar)=\sum_{k=0}^{\infty} \hbar^{k} \Phi_k(Q),
E_0(\hbar)=\sum_{k=0}^{\infty} \hbar^{k} E^{(k)},
E^{(0)}=n/2.
$$

This paper deals with the asymptotical behaviour of
$\Phi_k(Q)$ as $k$ is large and $Q$ is fixed. The problem of the high order
asymptotics of the eigenfunction perturbation theory when the expansion
is considered at fixed $x$ will be discussed in the next paper.

In one-dimensional quantum mechanics of a particle in a potential shown in
fig.1 which was considered in \cite{BLGZJ,ZJ} the exponent in
(\ref{D1*}) is expressed through the action $S$ on the trajectory starting
from zero and reaching at once the point $Q$ (solid line in fig.2).
This trajectory is the solution of the euclidean equation of motion which
can be obtained from the ordinary one by changing of the real time
 $t$ to the imaginary one $-i\tau$:
 \begin{equation}\label{I3}
 \frac{d^2}{d\tau^2} {\cal Q} = V^{'}({\cal Q}).
 \end{equation}
It appears that the asymptotics of the quantity $\Phi_k$ at larges $k$
can be expressed through the action $S_s$ on the other euclidean solution
which starts from zero, reaches the turning point and finishes also
at point $Q$ ( dashed line in fig.2 ).

As well as in one-dimensional case, in multidimensional case the main
contribution to the wave function is given by the classical euclidean
solution with the least action $S$, while high order asymptotics
of the semiclassical expansion is determined by the action $S_s$ on
another euclidean solution.

It is shown in this paper that high orders of
 $\Phi_k$ have the following asymptotic behaviour at larges $k$,
\begin{equation}
\Phi_k(Q) \sim \frac{(k-1)!}{(S_s(Q)-S(Q))^k},
\label{D4V}
\end{equation}
the pre-exponential factor is omited in this formula. From the expressions
obtained in this paper one can also find the asymptotic behaviour of the
ground state energy perturbation coefficients
 $E^{(k)}$ and carry out a check of the formulas obtained in
\cite{BLGZJ,BPZJ,RS} for various cases by the path integral technique.

Besides the potential shown in fig.1, other examples are also considered in
this paper. First, the case of a particle moving in a radial symmetric
potential in $n$-dimensional space is also considered. If the potential
depends on the distance of the origin as it is shown in fig.1, then classical
trajectories determining the asymptotics of the eigenfunction and the high
order behaviour of semiclassical expansion are analogous with the
trajectories shown in fig.2 by solid and dashed lines for one-dimensional
case.

Second, the example of the potential with degenerate minima (fig.3) is
also discussed. Classical solution determining the asymptotics of the
eigenfunction is analogous to shown in fig.2. On the other hand, not
classical solution but ''almost classical solution'' contributes to the
large order behaviour of semiclassical expansion. This ''almost solution''
starts at $\tau \rightarrow -\infty$ from the origin, then transits to
another minimum. The ''almost solution'' resides in this minimum for a
 long euclidean time and finally reaches the point $Q$.
As it was shown in \cite{BPZJ,ZJ}, if $Q=0$, then these ''almost solutions''
(instanton-anti-instanton pairs) contribute to large order
behaviour of perturbation theory for the ground state energy.

Finally, a particle on $n$-dimensional sphere in the external potential
depending only on one coordinate and having one minimum at the south pole
of the sphere which is a classical ground state (fig.4) is also considered in
this paper. One of the classical solutions shown in fig.4 by solid line
determines the behaviour of eigenfunction, another solution passing through
the north pole of the sphere determines the high order asymptotics of
semiclassical expansion. This solution is shown in fig.4 by dashed line.
An interesting feature of this asymptotics is the nullification of it at odd
$n$ which was found in ref.\cite{RS} for the ground state energy perturbation
theory. In this case  high order asymptotics is determined by the ''almost
solution'' which makes a loop around the sphere, resides in the south pole for
a long time and finally reachs the point $Q$.

\section*{2 Methods of finding the large order asymptotics of semiclassical
expansion}

Asymptotics of $\Phi_k$ can be found by various methods. First, the ground
state wave function can be expressed through the path integral
 (see, for example, \cite{HH})
over trajectories starting as  $\tau \rightarrow -\infty$ from zero
 and reaching the point $Q$ at $\tau=0$:
\begin{equation}
\int_{{\cal Q}(-\infty)=0,{\cal Q}(0)=Q} D{\cal Q} \exp(-\frac{1}{\hbar}
S[{\cal Q}])
\label{D3+}
\end{equation}
where  $S$ is the euclidean action of the theory and have the form:
 \begin{equation}\label{I2}
 S[{\cal Q}]=\int d\tau [\dot{\cal Q}^2/2+V({\cal Q})].
 \end{equation}
This integral can be evaluated as $\hbar \rightarrow 0$
by the Laplase method. As an exponential approximation this integral is
equal to
\begin{equation}
\exp\left(-\frac{1}{\hbar} \min_{{\cal Q}(-\infty)=0,{\cal Q}(0)=Q}  S[{\cal
Q}]\right),
\label{exp} \end{equation}
This formula coinsides with semiclassical tunnel asymptotics.
The pre-exponential factor and the corrections can be calculated with the
help of extraction of the factor
(\ref{exp}) from the path integral (\ref{D3+}), substitution
${\cal Q}={\cal Q}_{0}+q\sqrt{\hbar}$,
where  ${\cal Q}_{0}$ is the trajectory with the least action
in eq.(\ref{exp}), expansion of the integrand after this substitution
in terms of  $\sqrt{\hbar}$ and calculation of integrals of the products of
some polynomial in $q$ functions
by the Gauss exponent. Coefficients of odd powers in
 $\sqrt{\hbar}$ are equal to zero. Quantities  $\Phi_k$
can be expressed through the integrals
\begin{equation}
\Phi_k=\int Dq \oint_C \frac{dg}{2\pi i g^{2k+1}}
\exp\left(-\frac{1}{g^2}[S({\cal Q}_{0}+gq)-S({\cal Q}_{0})]\right),
\label{***} \end{equation}
where the contour $C$ dependind in general on $q$ runs around the origin
counterclockwise. After substitution
$
{\cal Q}_{0}+gq={\cal Q}, g=\nu/\sqrt{k}
$
 integrals (\ref{***}) take the form:
\begin{equation}\label{B3*}
\Phi_k=k^k \int \frac{D{\cal Q} d\nu}{2\pi i \nu^{2k+1}}
\exp\left(-k\left(
\frac{S[{\cal Q}]-S[{\cal Q}_{0}]}{\nu^2}+\ln\nu^2
\right)\right),
\end{equation}
and can be evaluated by saddle-point technique.

Consider saddle points of the exponent $({\cal Q}_{s},\nu_s)$.
It's variation in $\nu$ gives us the following condition,
$
\nu_{s}^2=S({\cal Q}_s)-S({\cal Q}_0),
$
variation in $\Phi$ leads to the condition
$
\delta S({\cal Q}_s)=0,
$
i.e. ${\cal Q}_s$ is the classical solution.
Therefore, asymptotics of the integral (\ref{B3*}) has the form  (\ref{D4V})
up to a pre-exponential factor, where
 $S_s(Q)=S[{\cal Q}_{s}],S(Q)=S[{\cal Q}_{0}]$.

Another technique to calculate the $\Phi_k$ high order asymptotics is based on
the direct analysis of the recursive relations for the $\Phi_k$ which can be
obtained from eq.(\ref{D2*})
 \begin{equation}
-\frac{1}{2}\Delta\Phi_{k-1}+\nabla S \nabla \Phi_k + \frac{1}{2} (\Delta S
-n) \Phi_k = \sum_{p=1}^{k} E^{(p)} \Phi_{k-p} \label{D5+}
 \end{equation}
Let us look for the large order asymptotics in a form
\begin{equation} \Phi_k \sim \frac{(k-1)!}{A(Q)^k}. \label{D5*}
\end{equation}
Substitution of formula (\ref{D5*}) to relations  (\ref{D5+})
gives the following equation for  $A$
 \begin{equation} \nabla A = -2 \nabla S \label{D5V} \end{equation}
 in the leading order. It has been supposed that high orders of
 $\Phi_k$
are growing faster than  $E^{(k)}$.
This assumption can be justified as follows. It follows from eq.(\ref{D5V})
that the value of the function $A$ in zero is more than in other
points of some vicinity of zero. Since the $E^{(k)}$ asymptotics is
distinguished from the $\Phi_k$ asymptotics only by a pre-exponential factor
and, therefore, has the form $\Gamma(k)/A(0)^k$, the right-hand side of
eq.(\ref{D5+}) can be neglected at $Q\ne 0$ because it is exponentially small.
On the other hand, when $A(0)-A(Q)\sim 1/k$, one can't ignore the
right-hand side,
so that calculations based on this neglection break down at
$Q\sim 1/\sqrt{k}$. Therefore, the asymptotics at these $Q$ and larges $k$
has another form discussed in section 5.

Section  3 contains the calculation of the pre-exponential factor in formula
 (\ref{D5*}) with the help of direct analysis of the recursive relations.
This factor is defined up to a multiplier, the function $A$
is determined up to an additive constant.

In section 3 there is also a consideration of the following interesting
method for calculating the pre-exponential factor and the corrections to
asymptotic formula. Let us look for the
$\Phi_{k}$ asymptotics as $k \rightarrow \infty$ when the corrections are
taken into account in a form
\begin{equation}
\Phi_k=B(Q,e^{-\partial/\partial k}) \frac{(k-1)!}{A(Q)^k}
\label{D6*}
\end{equation}
where $B$ is a sum of power functions
  of the operator $e^{-\partial/\partial k}$.
  Notice that the operator  $e^{-\partial/\partial k}$
playes the role of a small papameter in this case. Namely, it
transforms the sequence
 $(k-1)!/A^k$ to the sequence $(k-2)!/A^{k-1}$,
 the $k$-th order of the latter sequence is less than the $k$-th order of
the former one approximately in $k/A$ times.
Therefore, any asymptotics of the form
\begin{equation}
k^{\nu}\frac{(k-1)!}{A(Q)^k}(1+a_1/k+a_2/k^2+...)
\label{D6A}
\end{equation}
can be presented in a form
\begin{equation}
(b_0+b_1 e^{-\partial/\partial k}+b_2 e^{-2\partial/\partial k}+...)
 e^{\nu\partial/\partial k}\frac{(k-1)!}{A(Q)^k}
\label{D6B}
\end{equation}
i.e. in the form (\ref{D6*}). Coefficients $b$ in eq.(\ref{D6B}) can be
expressed through the coefficients $a$ in eq.(\ref{D6A}).
The $k$-th order of the ground state energy perturbation theory can be
presented in the form analogous with (\ref{D6*}),too.

These expansions for $E^{(k)}$ and $\Phi_k$ can be substituted to the
relations  (\ref{D5+}). Analysis of the obtained equation is analogous to
calculating of the corrections to semiclassical approximation. But in
this case the parameter of the expansion is not a number (Planck constant)
but an operator  $e^{-\partial/\partial k}$.
The general theory of the semiclassical expansion when
 $\hbar$ is an operator was developed in \cite{M1}.

Eq.(\ref{D5V}) determines the function $A$ up to an additive constant. For
determining it, one must analyse the behaviour of semiclassical approximation
near singular points: in one-dimensional case
it is the turning point that gives the singularity in eq.
(\ref{D4V}).

The behaviour of semiclassical expansion near the turning point which
determines the constants in formulas for the function $A$ and the
pre-exponential factor is analysed in section 4 for various types of
turning points.

It will be shown that the pre-exponential factor is divergent near the point
$Q=0$. The divergence is connected with the necessity of constructing another
asymptotics as $Q\sim 1/\sqrt{k}$, which is non-singular and allows us to
find large order asymptotics of the ground state energy perturbation
theory. Section 5 contains such derivation.
 The obtained results
coincide with the obtained one in refs.\cite{BLGZJ,BPZJ,RS}.

\section*{3 Analysis of the semiclassical expansion recursive relations}

In this section the derivation of the asymptotic formulas from the recursive
relations is considered in more details. Let us examine the following form
for the asymptotics of  $\Phi_k$ at larges $k$
\begin{equation}
 \Phi_k(Q) \sim \frac{\Gamma(k)k^{\nu}}{A(Q)^{k+\nu}}f(Q) \label{D14*}
\end{equation}
 and find conditions for the constant $\nu$  and functions  $A$, $f$.
Substitute the expression (\ref{D14*}) to the left-hand side of eq.(\ref{D5+})
and consider first terms of order $\Gamma(k)k^{\nu+1}$ and then terms of order
 $\Gamma(k)k^{\nu}$. To calculate the corrections to the asymptotics
 (\ref{D14*}) one can consider the following terms.

Let us use the relations

\begin{equation}
\nabla \Phi_k = -
\frac{\Gamma(k)k^{\nu}(k+\nu)}{A(Q)^{k+\nu+1}}
f(Q)\nabla A +
\frac{\Gamma(k)k^{\nu}}{A(Q)^{k+\nu}}
\nabla f + ... ,
\label{D14+}
\end{equation}
$$
\Delta \Phi_{k-1} =
\frac{\Gamma(k)k^{\nu}(k+\nu)}{A(Q)^{k+\nu+1}}
(\nabla A)^2 f(Q)
$$
\begin{equation}
 + \frac{\Gamma(k)k^{\nu}}{A(Q)^{k+\nu}}
[-f(Q)\Delta A - 2 \nabla A \nabla f]+...
\label{D14-}
\end{equation}
The right-hand side of eq.(\ref{D5+}) can be omitted if
 $ Q \ne 0$ because of the remark of section 2. Substitution of the relations
(\ref{D14+}) and (\ref{D14-}) to formula (\ref{D5+}) gives the equation
for  $A$ (\ref{D5V}) in a leading order, so that
$$
A(Q)=A_0-2S(Q),A_0=const.
$$
The next order gives us the following equation for  $f$,
\begin{equation}
\frac{1}{2} [ f\Delta A + 2 \nabla A \nabla f] + \nabla S \nabla f
+ \frac{1}{2} (\Delta S -n)f=0.
\label{*2}
\end{equation}

Separate now factor $\Phi_0$  from the function $f$ and denote
\begin{equation}
f/\Phi_0=X.
\label{D15f}
\end{equation}

It follows from the equations for $\Phi_0$ and $A$ that the function $X(Q)$
satisfies the condition,
\begin{equation}
\nabla S \nabla X = -nX
\label{D15P}
\end{equation}
and can be denote up to a multiplier. In one-dimensional case
 function $X$ has the form
\begin{equation}
X=c\exp(\int_{Q}^{Q_+} dQ/\sqrt{2V(Q)} )
\label{D15X}
\end{equation}
In section 4 the vicinity of the turning point is considered in more details
and constants
 $c$ and $\nu$ are found.
Other types of singular points are also considered.

Recursive relations for semiclassical expansion can be also investigated for
Hamiltonians with quantum corrections. For example, consider Hamiltonians
of the following type
\cite{RS},
\begin{equation}
H=-\frac{\hbar^2}{2}\frac{d^2}{dQ^2}
-\frac{n-1}{2Q}u(Q)\hbar^2\frac{d}{dQ}+V(Q),
\label{I12}
\end{equation}
where function $u(Q)$ is equal to 1 when  $Q=0$.
When $u=Qctg Q$, this Hamiltonian describes a particle on $n$-dimensional
sphere. If $u=1$ then eq.(\ref{I12}) corresponds to the $O(n)$-symmetrical
case.

Substitution of the formula
(\ref{D1*}) to the equation $H\psi=\hbar E_0\psi$
leads to the Hamilton-Jacobi equation coinciding to
 (\ref{*1}), while the equation for
$\Phi$ takes the more complicated form than  (\ref{D2*}):
$$
-\frac{\hbar}{2}\Phi^{''} + S^{'}\Phi^{'} + \frac{1}{2} (S^{''}-n)\Phi
+\frac{n-1}{2Q}uS^{'}\Phi - \hbar \frac{n-1}{2Q} u \Phi^{'} =
(E_0-n/2)\Phi.
$$
Recursive relations for the perturbation series coefficients in $\hbar$
can be also obtained from the equation for $\Phi$.

Asymptotics of the $\Phi_k$ at larges $k$ can be also looked for in a form
(\ref{D14*}). Substitution of this formula to the recursive relations
gives the equation for
$f$ distinguished from  (\ref{*2}). But after extracting  the factor
 $\Phi_0$ (eq.(\ref{D15f})) the relation for the function  $X$
takes, nevertheless, the form (\ref{D15P}). Its solution has the form
$$
X=c\exp\left(n\int_{Q}^{Q_+} \frac{dQ}{\sqrt{2V(Q)}}\right).
$$

One can find the corrections to the asymptotic formula
(\ref{D14*}) by evaluating the following terms in eqs.(\ref{D14+}),
(\ref{D14-}), substituting it to the left-hand side of the eq.
(\ref{D5+}) and setting the coefficients of the corresponding orders in
 $k$ equal to zero. Nevertheless, in this section another technique
to find the corrections is examined. This method illustrates the
analogy between the corrections to semiclassical approximation and
the corrections to high order asymptotics.

As it has been mentioned in section 2, let us search for the asymptotics
of the $\Phi_k$ in a form (\ref{D6*}). When $Q \ne 0$, one can present the
eq.(\ref{D5+}) in a form
$$
-\frac{1}{2}
e^{-\partial /\partial k}
\Delta \Phi_k + \nabla S \nabla \Phi_k +\frac{1}{2}(\Delta S - n)\Phi_k=
$$
\begin{equation}
=(E^{(1)} e^{-\partial /\partial k} +E^{(2)} e^{-2\partial /\partial k}
+...)\Phi_k,
\label{D16+}
\end{equation}
where the exponentially small terms associated with the ground state energy
asymptotics are omitted.

As it has been shown in section 2, the operator
$e^{-\partial /\partial k}$
plays a role of a small parameter analogous to
$\hbar$ in a case of semiclassical expansion.

Notice that the commutation rule of the derivation operator and
the function
$\Gamma(k)A(Q)^{-k}$ can be presented in a form
\begin{equation}
\frac{\partial}{\partial Q} \frac{\Gamma(k)}{A(Q)^k} = -
e^{-\partial /\partial k}
\frac{\Gamma(k)}{A(Q)^k}\frac{\partial A}{\partial Q} +
\frac{\Gamma(k)}{A(Q)^k}
\frac{\partial}{\partial Q}
\label{D16*}
\end{equation}
analogous to the commutation formula of the derivation operator with the
exponent
$$
\frac{\partial}{\partial Q}\exp(-A/\hbar)
=-\frac{1}{\hbar}\exp(-A/\hbar)\frac{\partial A}{\partial Q} +
\exp(-A/\hbar)
\frac{\partial}{\partial Q},
$$
where a number $\hbar$ is substituted by the operator
$e^{-\partial /\partial k},$
while the function of $\hbar$, $\exp(-A/\hbar)$, is changed by the set
of numbers,
 $\Gamma(k)A^{-k}$. Application of eq.(\ref{D16*})
 to  eq.(\ref{D16+}) leads to the equation for the function
 $B$ presented in formula
 (\ref{D6*}),
 $$
 -\frac
{e^{-\partial /\partial k}}{2}
(\nabla - e^{\partial /\partial k} \nabla A)^2 B +
\nabla S
(\nabla - e^{\partial /\partial k} \nabla A) B
$$
$$
+\frac{1}{2}(\Delta S - n)B=
(e^{-\partial /\partial k}E^{(1)} +
e^{-2\partial /\partial k}E^{(2)} + ...)B
$$
which is analogous to the equation for
$\Phi$ (\ref{D2*}). Let us set the coefficients of each order in
$e^{-\partial /\partial k}$
equal to zero. We obtain the equations derived earlier and the following
formula for the corrections,
$$
B(e^{-\partial /\partial k},Q)=B_0(Q)\sum_{l=0}^{\infty} F_l(Q)
e^{-l\partial /\partial k},
$$
$$
-\frac{1}{2B_0}
\Delta(F_{l-1}B_0)+\nabla S \nabla F_l = \sum_{s=0}^{l-1} E^{(l-s)}F_s
$$
These equations define the function $B$ up to the factor which does not
depend on $Q$ but depends on
$e^{-\partial /\partial k}$.
Let us consider the singular points and find this factor.

\section*{4 The behaviour of the asymptotics near singular points}

As it has been noticed in the  previous section, one must consider the
behaviour of the recursive relations near singular points in order to
find constants $c,A_0,\nu$. This problem is examined in this section. The main
idea of the consideration is the following. The potential near singular point
can be approximated by a linear, quadratic or another function depending on the
type of a singular point. All the orders of semiclassical expansion can be
evaluated exactly for this approximate potential, so one can find unknown
constants by comparing the ''exact'' results with the asymptotics discussed in
section 3.

In this section three types of one-dimensional singularities are discussed:

i) ordinary turning point (fig.1);

ii) potentials with degenerate minima (fig.3);

iii) singular quantum correction to the Hamiltonian \cite{RS}.

Consider these cases more preceisely.

\subsection*{4.1 Ordinary turning point}

In this subsection the case of one-dimensional quantum mechanics of a particle
in the potential shown in fig.1 is considered. Let us discuss the behaviour of
semiclassical expansion for the ground state wave function in the vicinity of
point $Q_+$.The potential $V$ can be approximated by a linear function,
\begin{equation}
V \sim a\xi, \xi=Q_+-Q,
\label{D19+}
\end{equation}
The function $A$ satisfying the equation (\ref{D5V}) is approximately equal to
$$
A=A_0-2S_++\frac{4}{3}\sqrt{2a}\xi^{3/2},
$$
where $S_+$ is the action
$
S=\int_{0}^{Q_+} \sqrt{2V(Q)} dQ
$
 at the turning point $Q_+$. As the duration of motion from the point
 $Q$ to the turning point has the form
$
\int_Q^{Q_+} \frac{dQ}{\sqrt{2V(Q)}} = \sqrt{2\xi/a},
$
eq.(\ref{D14*}) implies that the high orders of semiclassical expansion
behave as follows,
 \begin{equation} \Phi_k \sim \frac{\Gamma(k)
k^{\nu}ce^{\sqrt{2\xi/a}}}
{(A_0-2S_++\frac{4}{3}\sqrt{2a}\xi^{3/2})^{k+\nu}} \Phi_0.
\label{D19*}
\end{equation}
Let us obtain the asymptotics (\ref{D19*}) in another way and find the
coefficients $c,\nu,A_0$.

First of all, notice that the semiclassical expansion for the wave function in
the case of a potential (\ref{D19+}) can be obtained as follows.
Formula (\ref{D16+}) implies that the function
$
 \psi=\sum \Phi_k \hbar^k e^{-S/\hbar}
$
approximately satisfies as an asymptotic series the equation
$$
(-\frac{\hbar^2}{2}\frac{d^2}{d\xi^2} + a\xi) \psi(\xi)=0
$$
The expansion in powers of $\hbar$ of the growing at large $\xi$
solution can be obtained from the following expression,
\begin{equation}
\psi(\xi)=\int dp
 \exp(-\frac{1}{\hbar}(p\xi-\frac{p^3}{6a})
\label{I10}
\end{equation}
where the integral is taken over any sufficiently small region containitg the
minimum of the exponent, $p_0=-\sqrt{2a\xi}$.
It follows from the integral presentation (\ref{I10}) that the
 $\Phi_k$ asymptotics at larges $k$ has the form
$$
\Phi_k \sim \Phi_0 \frac{1}{2\pi} \frac{\Gamma(k)e^{\sqrt{2\xi/a}}}
{(\frac{4}{3}\xi \sqrt{2a \xi})^k},
$$
coinciding with eq.(\ref{D19*}) when
\begin{equation}
\nu=0,A_0=2S_+,c=1/(2\pi).
\label{I11}
\end{equation}

Substitution of these constants to the formulas obtained in section 3 gives
us the following asymptotics,
\begin{equation}
\Phi_k(Q) \sim \frac{(k-1)!\Phi_0(Q)}{2\pi (S_s(Q)-S(Q))^k}
\exp(\int_{Q}^{Q_+}\frac{dQ}{\sqrt{2V(Q)}}).
\label{D13*}
\end{equation}

\subsection*{4.2 Potential with degenerate minima }

Consider now the case of the double-well potential shown in fig.3. Namely,
let the potential $V$ have the minimum in the point $Q_+$,
 besides the minimum
at zero and let $V(Q_+)$ equal to zero.In the vicinity of the point
$Q_+$ the potential can be approximated by a quadratic function,
$$
V \sim \omega^2 \xi^2/2, \xi=Q_+-Q.
$$
Action  $S$  has the following form in this approximation,
$$
S=S_+-\omega \xi^2/2
$$
The duration of motion from  point  $Q$ to point $Q_+$ is infinite in this
case, contrary to the previous subsection. Therefore, the solution $X$ to
eq. (\ref{D15P}) has the more complicated form than (\ref{D15X}).
Namely,
 \begin{equation} X=c(Q_+-Q)^{1/\omega}\exp \int_{Q}^{Q_+} dQ\left[
\frac{1}{\sqrt{2V(Q)}} - \frac{1}{\omega (Q_+ -Q)}
\right],
\label{D21*}
\end{equation}
in this formula singular contribution is substracted from the exponent in
eq.(\ref{D15X}) and the normalizing factor is redefined. One can show by
the explicit calculation that the function
(\ref{D21*}) really satisfies  eq.(\ref{D15P}).

In the vicinity of the point $Q_+$ the asymptotic formula (\ref{D14*})
takes the form
\begin{equation}
\Phi_k \sim \frac{\Gamma(k) k^{\nu}c \xi^{1/\omega}}
{(A_0-2S_++\omega \xi^2)^{k+\nu}} \Phi_0
\label{D21+}
\end{equation}

On the other hand, recursive relations have the following form in the
quadratic approximation
\begin{equation}
-\frac{1}{2}\frac{d^2\Phi_{k-1}}{d\xi^2}-\omega \xi \Phi_0
\left(\frac{\Phi_k}{\Phi_0}\right)^{'} = 0
\label{D22*}
\end{equation}
and can be solved exactly,
\begin{equation}
\Phi_k=\frac{c_k}{\xi^{2k}} \Phi_0,\Phi_0=\xi^{-\frac{1+\omega}{2\omega}}
\label{D22F}
\end{equation}
where numerical coefficients $c_k$ have the form
$$
c_k=\frac{\Gamma(2k+\frac{1+\omega}{2\omega})}
{(4\omega)^k \Gamma(k+1) \Gamma(\frac{1+\omega}{2\omega})}
$$
Making use of the Stirling formula for the Gamma-function, one can obtain that
at large  $k$ the  $c_k$ asymptotics can be written in a form
\begin{equation}
c_k \sim \frac{(k-1)!}{\omega^k \sqrt{2\pi}} \frac{(2k)^{1/\omega}}
{\Gamma(\frac{1+\omega}{2\omega})}
\label{D22C}
\end{equation}
The constants in the $\Phi_k$ asymptotics can be found by comparing the
formulas (\ref{D22C}),(\ref{D22F}) with the formula (\ref{D21+}).
The constants are the following,
\begin{equation}
c=\frac{(2\omega)^{\frac{1}{2\omega}}}
{\sqrt{2\pi}\Gamma(\frac{1+\omega}{2\omega})},
\nu=\frac{1}{2\omega},A_0=2S_+.
\label{D22+}
\end{equation}
It follows from the formulas (\ref{D14*}),(\ref{D15f}),
(\ref{D22+}),(\ref{D21*}) that the
 $k$-th order of the ground state wave function semiclassical expansion
has the following asymptotics as
 $k \rightarrow \infty$ in the case of a degenerate minima potential,
$$
\Phi_k \sim \Phi_0 \frac{\Gamma(k)k^{1/2\omega}}{(2S_+-2S(Q))^{k+1/2\omega}}
\frac{(2\omega)^{\frac{1}{2\omega}}(Q_+-Q)^{1/\omega}}
{\sqrt{2\pi}\Gamma(\frac{1+\omega}{2\omega})}
 $$
\begin{equation}
\times\exp \int_{Q}^{Q_+} \left[
\frac{1}{\sqrt{2V(Q)}} - \frac{1}{\omega (Q_+ -Q)}
\right]
\label{D23AS}
\end{equation}

\subsection*{4.3 Singular quantum correction to the Hamiltonian}

Consider the Hamiltonians presented as a sum of a classical (non-singular)
Hamiltonian and a quantum correction to it which is singular,
\begin{equation}
H=-\frac{\hbar^2}{2}\frac{d^2}{dQ^2} + V(\cos Q) -
\frac{\hbar^2}{2}(n-1) \frac{\cos Q}{\sin Q}\frac{d}{dQ}
\label{C12*}
\end{equation}
Evaluation of the high order asymptotics for the ground state energy
perturbation theory in this case was considered in \cite{RS}.
The $\Phi_k$ asymptotics in this case has the form
(\ref{D14*}), the function $X$ satisfies the equation (\ref{D15P})
at non-singular points  ($Q \in (0,\pi) $)
and, therefore, has the form (\ref{D15X}), where $Q_+=\pi$.

When one study the vicinity of the singular point in this case, one must take
into account not only classical part of the Hamiltonian but also the
quantum correction to it because it is singular, as oppose to the previous
case. The classical Hamiltonian can be approximated by the Hamiltonian of
a free particle,
\begin{equation}
-\frac{\hbar^2}{2}\frac{d^2}{d\xi^2} + E, \xi=\pi-Q,
\label{D23A}
\end{equation}
while the quantum correction is approximately equal to the following
operator,
\begin{equation}
-\frac{\hbar^2}{2}(n-1)\frac{1}{\xi}\frac{d}{d\xi}.
\label{D23B}
\end{equation}

In the vicinity of the singular point the asymptotic formula (\ref{D14*})
takes the following approximate form,
\begin{equation}
\Phi_k \sim \frac{\Gamma(k) k^{\nu}c e^{\xi/\sqrt{2E}} }
{(A_0-2S_++2\sqrt{2E}\xi)^{k+\nu}} \Phi_0
\label{D24V}
\end{equation}
The constants can be found by comparing with this formula. The function
$
\psi=\sum \hbar^k \Phi_k e^{-S/\hbar}
$
satisfies the equation
$$
(-\frac{\hbar^2}{2}\frac{d^2}{d\xi^2} + E
-\frac{\hbar^2}{2}(n-1)\frac{1}{\xi}\frac{d}{d\xi})\psi(\xi)=0.
$$
Its solution growing at infinity is considered. It can be expressed through
the Infeld function $I_{n/2-1}$ (the Bessel function of a purely imaginary
argument),
$$
\psi(\xi)=\xi^{1-n/2} I_{n/2-1}[\xi\sqrt{2E}/\hbar].
$$
Semiclassical expansion of the fnction $\psi$ corresponds to the
Infeld function expansion at large arguments. As this asymptotic series
breaks for the Infeld function of the half-integer order, semiclassical
expansion also breaks, so in the approximation
 (\ref{D23A}),(\ref{D23B}) all the  orders of $\Phi_k$
begining from some $k$ are equal to zero. Therefore, singular
point
 $Q_+=\pi$ does not contribute to the high order asymptotics of semiclassical
expansion at integer odd $n$. This confirms the assumption of \cite{RS}.

If $n$ is not equal to an odd integer number, the Infeld function expansion
coefficients
$$
I_{\nu}(x)=\frac{1}{\sqrt{2\pi x}} e^x (1+c_1/x+c_2/x^2+...)
$$
have the form \cite{RG}:
$$
c_k=\frac{\cos \pi\nu}{\pi} \frac{\Gamma(k-\nu+1/2)\Gamma(k+\nu+1/2)}{2^kk!}.
$$
and the following asymptotic behaviour at larges $k$
$$
c_k \sim \frac{\Gamma(k)}{2^k} \frac{\cos \pi \nu}{\pi}.
$$
Therefore, the semiclassical expansion of the function
 $\psi$ behaves at high orders as follows,
$$
\Phi_k \sim \Phi_0 \frac{\Gamma(k)}{(2\xi)^k(2E)^{k/2}}
\frac{\cos\pi(n/2-1)}{\pi}
$$
Therefore, the constants $A_0,c,\nu$ in the asymptotic formula for
 $\Phi_k$ have the form,
$$
A_0=2S_+,\nu=0,c=-\frac{\cos (\pi n/2)}{\pi}
$$
Thus, the following expression for the  $\Phi_k$ asymptotics is obtained,
\begin{equation}
\Phi_k \sim \Phi_0 \frac{\Gamma(k)}{(2S_+-2S(Q))^{k}}
(-\frac{1}{\pi} \cos\pi n/2)
\exp \int_{Q}^{Q_+}dQ \frac{1}{\sqrt{2V(\cos Q)}}
\label{D25AS}
\end{equation}

\section*{5 The behaviour of the asymptotics in the vicinity of the minimum
and high order behaviour of the ground state energy perturbation theory}

The asymptotics obtained in the previous section for various cases have
singularities in the origin. Namely, in one-dimensional case at small
 $Q$ the asymptotic formula for $\Phi_k$ is approximately equal to
 \begin{equation} \Phi_k(Q) \sim
\frac{B}{|Q|}\frac{(k-1)!k^{\nu}}{(A_0-Q^2)^{k+\nu}}, \label{D26*}
\end{equation}
(the factor $B$ depends on the potential), i.e. the pre-exponential factor
diverges as $1/Q$. In the $n$-dimensional case there is a divergence as
 $1/|Q|^n$.

On the other hand, in the each order of semiclassical expansion the wave
function is finite at zero. It is the behaviour of the eigenfunction
near zero that allows us to find an asymptotic behaviour of the
perturbation theory for eigenvalues.

It occurs, nevertheless, that the $\Phi_k(Q)$ asymptotics considered not
under the conditions $Q=const,k \rightarrow \infty$
but under other conditions
\begin{equation}
k \rightarrow \infty, Q\sqrt{k} \rightarrow y=const,
\label{D26+}
\end{equation}
is non-singular at zero argument.
 The knowledge of this asymptotics enables us
to find high order behaviour of the eigenvalue perturbation theory.

It is convenient to change the variable $y=z\sqrt{A_0}$.
Let us search for the  $\Phi_k$ asymptotics under the conditions (\ref{D26+})
in a form:
\begin{equation}
\frac{(k-1)!k^{\nu+1/2}}{A_{0}^{k+\nu+1/2}} g(z),
\label{D27+}
\end{equation}
while the  $E^{(k)}$ asymptotics is seeking in a form:
 \begin{equation} E^{(k)} \sim
\frac{(k-1)!k^{\nu+1/2}}{A_{0}^{k+\nu+1/2}} {\cal E} \label{D27B}
\end{equation}
Consider the substitution of the formulas (\ref{D27+}) and (\ref{D27B})
to the equation (\ref{D5+}). As the asymptotics is considered near the point
$Q=0$, one cannot omit the right-hand side of the equation
 (\ref{D5+}). At larges $k$ the main contribution to the sum in the right-hand
side of the formula (\ref{D5+}) is given by the $k$-th term equal to
$E^{(k)}$ because $\Phi_0=1$ in the origin. Notice also that
 $ \Delta S(0) = n$.  Therefore, at larges $k$ eq.(\ref{D5+}) takes the
following form in the leading order,
\begin{equation}
-\frac{1}{2}\frac{d^2g}{dz^2}+z\frac{dg}{dz}={\cal E}.
\label{D27A}
\end{equation}
Find now the connection between the quantity $\cal E$ defining the large orders
of the ground state energy perturbation theory and the factor
$B$ obtained from the wave function high order asymptotics.

Consider first the case of a symmetric potential. Formula
(\ref{D26*}) takes then place at positive $Q$ as well as at negative $Q$.
At larges $z$ an asymptotic formula for
$\Phi_k(z\sqrt{A_0/k})$ must transform to the asymptotics (\ref{D26*}) where
the following change, $Q=z\sqrt{A_0/k}$, is made. Let us substract the factor
$A_0^{-k}$ from eq.(\ref{D26*}) and make a limit (\ref{D26+}).
Let us also take into account the relation,
$
(1-z^2/k)^{k+\nu} \rightarrow e^{-z^2}.
$
The following boundary condition for function  $g$ at $+\infty$, as well as at
 $-\infty$ are obtained from it:
\begin{equation}
g(z) \sim \frac{B }{|z|}e^{z^2}.
\label{D27-}
\end{equation}
On the other hand, eq.(\ref{D27A}) can be solved exactly for the derivative
$g^{'}(z)$. Comparing this solution with the asymptotics for
 $g^{'}(z)$ obtained from eq.(\ref{D27-}), one can find that:
$$
{\cal E}=-\frac{2B}{\sqrt{\pi}}.
 $$
 The constant
 $B$ and, therefore, the quantity  $\cal E$ can be found for various cases
considered in the previous section. Thus, ground state energy perturbation
theory has the following large order asymptotic behaviour,
$$
E^{(k)} \simeq -\frac{k!k^{-1/2}}{\pi^{3/2}S_{B}^{k+1/2}}
Q_{+}\exp\int_{0}^{Q_{+}} dr[1/\sqrt{2V(r)}-1/r],
$$
in the case of ordinary turning point discussed in subsection 4.1, while in
the case of the potential with degenerate minima treated in subsection 4.2
the asymptotics is as follows,
$$
E^{(k)} \sim -\frac{2k!\sqrt{\omega}}{\pi A_{0}^{k+1}}
 \left(\frac{2k\omega}{A_0}\right)^{\frac{1-\omega}{2\omega}}
\frac{Q_{+}^{1/\omega+1}}{\Gamma(\frac{1+\omega}{2\omega})}
$$
$$
\times \exp \int_{0}^{Q_+} dQ \left(\frac{1}{\sqrt{2V(Q)}}-
\frac{1}{\omega(Q_+-Q)}-\frac{1}{Q}\right).
$$
These formulas are in agreement with the results obtained in
  \cite{BLGZJ,BPZJ}.

When the potential is not symmetric, the function  $A$ in the asymptotics
 (\ref{D5*}) has discontinuity at zero argument. For the definiteness, assume
that the least value of $A$ corresponds to positive $Q$.
Then the boundary condition at $+\infty$ has as in the considered case the form
 (\ref{D27-}),
while the condition at $-\infty$ is the boundness of $g$.
It follows from solving  eq.(\ref{D27A}) with these boundary conditions that
the quantity
 $\cal E$ is in  2 times lesser than in the previous case:
$
{\cal E}=-\frac{B}{\sqrt{\pi}}.
$
This result has the following interpretation in terms of the path integral
approach. When the potential is even, there are two classical solutions
symmetric under the change $Q$ to $-Q$ which contribute to the asymptotics,
as opposed to  non-even potentials.

Consider now the case of Hamiltonians
 (\ref{I12}). The asymptotic formula for
$\Phi_k$ transforms at small  $Q$ to the following expression,
$$
\frac{(k-1)!}{(A_0-Q^2)^k}\frac{B}{|Q|^n}.
$$

Let us seek for the asymptotics  of $\Phi_k$ under the conditions (\ref{D26*})
in a form,
\begin{equation}
\frac{(k-1)!k^{n/2}}{A_{0}^{k+n/2}} g(Q\sqrt{k/A_0}),
\label{I14}
\end{equation}
analogous to one-dimensional case, where function $g(z)$ satisfies the
following boundary conditions at larges $z$,
\begin{equation}
g(z)\sim B e^{z^2}/z^n.
\label{I16}
\end{equation}
Let us look for the asymptotics of $E^{(k)}$ in a form,
\begin{equation}
E^{(k)} \sim \frac{(k-1)!k^{n/2}}{A_{0}^{k+n/2}} {\cal E}.
\label{I15}
\end{equation}
Substitution of the formulas (\ref{I14}) and (\ref{I15}) to the
recursive relations of semiclassical expansion gives us the following
equation for $g$,
$$
-\frac{1}{2}g^{''}(z) - \frac{n-1}{2z}g^{'}(z) + zg^{'}(z) = {\cal E}.
$$
Its solution which is an even function of $z$ and equal to zero at
$z=0$ have the form
$$
g(z)=\int_{0}^{z} \frac{dz}{z^{n-1}} e^{z^2}
\int_{0}^{z} [-2{\cal E}z^{'n-1}e^{-z^{'2}}]dz^{'}.
$$
Asymptotic behaviour of this formula at larges $z$ is in agreement with the
expression (\ref{I16}) when
$$
{\cal E} = -\frac{2B}{\Gamma(n/2)}.
$$
Substitution of the expression for $B$ to this formula leads to the
asymptotics (\ref{I15}) coinciding in the case of the
$O(n)$-symmetric systems with the results of
 \cite{BLGZJ},
\begin{equation}
E^{(k)} \simeq -\frac{k!k^{n/2-1}}{\pi\Gamma(n/2)S_{B}^{k+n/2}}
(Q_{+}\exp\int_{0}^{Q_{+}} dr[1/\sqrt{2V(r)}-1/r])^{n}, \label{C35+}
\end{equation}
 and to the asymptotics
 \begin{equation} E^{(k)} \simeq \frac{k!k^{n/2-1}}{S_{SI}^{k+n/2}}\cos(\pi
n/2)
\frac{2\pi^{n-1}}{\Gamma(n/2)}\exp\left(n\int_{0}^{\pi}d\theta \left(\frac{1}{
\sqrt{2V(\cos\theta)}}-\frac{1}{\pi-\theta}\right)\right)
\label{C24*}\end{equation}
in the case of the Hamiltonians (\ref{C12*}).
The asymptotics (\ref{C24*}) is in agreement with \cite{RS}.

\section*{6 Conclusions}

In this paper large order asymptotics of the tunnel semiclassical expansion
for quantum mechanical systems is considered. As usual, the dependence of the
Hamiltonian on the semiclassical expansion parameter (Planck constant) has
the form (\ref{*3}). The results related to the high order asymptotics for the
ground state energy are in agreement with the papers
\cite{BW1,BW2,BLGZJ,BPZJ,ZJ}. Contrary to them, this paper deals with the study
of large order asymptotics not only for eigenvalues but also for
eigenfunctions.
Therefore, one can examine this problem not only by the path integral
approach but also by the direct analysis of the semiclassical expansion
recursive relations.

When one constructs the asymptotics, an important role is played by
classical euclidean solutions starting from the origin and finishing
at the point $Q$. These solutions give an exponentially small contribution
to the wave function in comparison with the contribution of another
solution satisfying the same boundary conditions. An interesting feature of
the constructed asymptotics for the semiclassical expansion large orders
is the divergence of the pre-exponential factor near zero value
of argument. This
difficulty is resolved by the investigation of the asymptotics as
$Q\sim 1/\sqrt{k}$. Another interesting feature of the obtained asymptotic
formula is the following.In one-dimensional case the most essential
growth of the asymptotics takes place near the turning point. It is the
comparison
of the asymptotics near this point with the exact coefficients of the
expansion for the linear potential that allows
 us to find the unknown constants in
the asymptotic formula. Analogous tecnique is applicable to other cases of
the singular points.

Thus, the approach suggested in this paper enables us to obtain the
high order asymptotics of semiclassical expansion both for eigenvalues and
eigenfunctions. Although only the case of the ground state have been also
examined, an analogous treatment can be also applicable to the excited
states.

Probably,one can generalize the discussed method to quantum field theory and
find the asymptotic behaviour of a sum of Feynman diagramms with $N$ external
lines and $k$ loops as both $N$ and $k$ tend to infinity. The considered
technique can be also useful in the case of instantons.

The author is indebted to V.A.Rubakov for helpful discussions.
This work is supported in part by ISF, grant \# MKT000.

{\bf Figure captions}

{\bf Fig.1.} Example of a one-dimensional potential $V(Q)$. This
function has a local minimum as its argument equals to zero,
$V(Q)\sim Q^2/2$, is positive when $0<Q<Q_+$ and negative as $Q_+<Q$.

{\bf Fig.2.} Solid line: trajectory giving the main contribution
to the ground state wave function. Dashed line: trajectory determining
the large order behaviour of semiclassical expansion.

{\bf Fig.3.} Potential $V(Q)$ with degenerate minima. This function
has two minima as $Q=0$ and $Q=Q_+$, $V(Q)=V(Q_+)=0$, and is positive
as $Q\ne 0$,$Q\ne Q_+$.

{\bf Fig.4.} A particle moving on a two-dimensional sphere. Solid and
dashed lines are, as well as in fig.2, classical euclidean solutions
contributing to wave functions and to large order asymptotics.


\begin{thebibliography}{99}
\bibitem{D} F.J.Dyson  {\em Phys. Rev.} {\bf 85} (1952) 631.
\bibitem{L}  L.N.Lipatov. {\em Sov. Phys. JETP} {\bf 45} (1977) 216.
\bibitem{BLGZJ} E.Br\'{e}zin, J.C.Le Guoillou, J.Zinn-Justin {\em Phys.Rev.}
                  {\bf D15} (1977) 1558.
\bibitem{BPZJ} E.Br\'{e}zin, G.Parisi, J.Zinn-Justin {\em Phys.Rev.} {\bf D16}
 (1977) 408.
\bibitem{ZJ} J.Zinn-Justin {\em Phys.Rep.}{\bf 70} (1981) 109.
\bibitem{RS} V.A.Rubakov, O.Yu.Shvedov. {\em Nucl.Phys.} {\bf B434} (1995) 245.
\bibitem{LL} L.D.Landau, E.M.Lifshitz. Quantum Mechanics. Non-relativistic
             Theory. Moscow, Nauka, 1989.
\bibitem{M1} V.P.Maslov. Asymptotic methods and perturbation theory.
      Moscow,Nauka,1988.
\bibitem{HH}  J.Hartle, S.Hawking {\em Phys.Rev.}{ \bf D28} (1983) 2960.
\bibitem{RG} I.S.Gradshtein, I.M.Ryzhik. Table of Integrals,Series and
Products. Academic Press, N.-Y.,1980.
\bibitem{BW1} C.M.Bender, T.T.Wu {\em Phys.Rev.} {\bf 184} (1969) 1231.
\bibitem{BW2} C.M.Bender, T.T.Wu {\em Phys.Rev.} {\bf D7} (1973) 1620.

\end{thebibliography}
\end{document}